\documentstyle[11pt, aas2pp4]{article}
\begin{document}
\title{A Search for Stellar Obscuration Events due to Dark Clouds}

\author{A.~J. Drake\altaffilmark{1,2,3}, K.~H. Cook\altaffilmark{1}}

\altaffiltext{1}{Lawrence Livermore National Laboratory, 7000 East Ave, Livermore, CA 94550}
\altaffiltext{2}{Dept. of Astrophysical Sciences, Princeton University, Princeton, NJ 08544}
\altaffiltext{3}{Depto. de Astronomia, P. Universidad Catolica, Casilla 104, Santiago 22, Chile}

\begin{abstract}

  The recent detections of a large population of faint submillimetre sources, an
  excess halo $\gamma$-ray background, and the extreme scattering events observed
  for extragalactic radio sources have been explained as being due to
  baryonic dark matter in the form of small, dark, gas clouds.
  In this paper we present the results of a search for the transient stellar
  obscurations such clouds are expected to cause. 
  We examine the Macho project light curves of $48 \times 10^{6}$ stars
  toward the Galactic bulge, LMC and SMC for the presence of dark cloud
  extinction events.  We find no evidence for the existence of a population
  of dark gas clouds with A$_V > 0.2$ and masses between $\sim 10^{-4}$ and
  $\rm 10^{-2} M_{\sun}$ in the Galactic disk or halo. However, it is
  possible that such dark cloud populations could exist if they are
  clustered in regions away from the observed lines of sight.

\end{abstract}

\section{\sc Introduction}

It has been proposed that a large fraction of the baryonic dark matter in
the halo of our Galaxy could be in the form of cold, dense gas clouds (de
Paolis et al.~1995, Gerhard \& Silk 1996).  Such clouds would be extremely
difficult to detect directly with traditional means due to their dark
nature.  However, the extreme scattering events observed by Fiedler (1987)
have been given as evidence for AU-sized, planetary-mass gas clouds
(Henriksen \& Widrow 1995, Walker \& Wardle 1998). %$\rm \sim 10^-4M_{\sun}$
Further evidence for the existence of such clouds comes from the detections
of large populations of faint sub-mm sources detected by SCUBA
(Submillimetre Common User Bolometric Array) (Lawrence 2001).  Yet more
evidence for such a population comes from recent EGRET (Energetic Gamma Ray
Explorer Telescope) results which show the presence of a background
$\gamma$-ray emission from the Galactic halo.  Kalberla, Schekinov and
Dettmar (1999) have suggested that these results can be explained as due to
the interaction of high energy cosmic rays with dense $\rm H_{2}$ clumps
with masses $\rm \sim10^{-3}M_{\sun}$ and radii $\sim 6$AU.  These
detections and prior observational and dynamical arguments have lead to
further suggestions about how a halo dark cloud population could be
discovered (Kerins, Binney \& Silk 2002, hereafter KBS; Kamaya and Silk 2002).
Although most models place the cloud population within the halo, Pfenniger and Combes
(1994) suggested that such clouds are a natural part of the cold ISM and
would exist concomitant with the thin HI disk.

For dark clouds to have survived until the present day, they must have a mass
which is sustainable over a Hubble time.  Within the Galaxy the lower mass
limit is $10^{-6} M_{\sun}$ because smaller masses will evaporate due to
cosmic ray heating (Wardle \& Walker 1999).  It is also expected that these
objects will contain some metals since it has been suggested that clouds
with primordial composition can not cool below 100K (Murray \& Lin 1990).
Any population of gas clouds must be much cooler than this to have escaped
detection in previous surveys (Lawrence 2001).  The presence of a small
amount of metals
can assist the cooling of gas clouds (Gerhard \& Silk 1996).  %$\sim 5 - 20$K
It has been suggested that, even if the clouds were dust-free, they could be
stabilized against gravitational collapse by cosmic ray heating (Sciama 2000).
However, if the SCUBA sources are due to clouds they must contain some dust
to exhibit continuous emission in the sub-mm band (Lawrence 2001). The SCUBA
results led to likely cloud masses of $\rm 10^{-4} - 2\times10^{-2}M_{\sun}$ with temperatures $<
18$K.  The virial radius of a cloud with $T = 18$K and $\rm M =
10^{-4}M_{\sun}$ is $0.39$AU, and for a cloud with T $= 10$K and $\rm M =
10^{-2}M_{\sun}$ the virial radius is 70AU.  These radii set likely-limits
on the sizes of a gas cloud population.  Further limits on clouds
sizes and masses come from constraints on gaseous lensing events.  Raficov
and Draine (2001) note that no gaseous lensing events have been observed
toward the LMC by the MACHO project, while many events are expected for a
significant population of compact gas clouds.  Massive compact objects such
as clouds, stars, or other objects will lense stars as they pass through our
line-of-sight of the stars.  Since lensing is not observed any clouds present 
must exceed the projected Einstein radius.
The analysis of Raficov and Draine (2001) suggests that cloud must have
radii from 2 to 300AU, for cloud masses between M $= 10^{-4}$ and $\rm
2\times10^{-2}M_{\sun}$ for temperatures of $10$ to $100$K.  Another
constraint on the dark clouds comes from the EGRET results of Kalberla et
al.~(1999).  They find that the $\gamma$-ray results are consistent with
$\rm H_{2}$ clumps with masses $\rm 0.3-3\times10^{-3}M_{\sun}$ and sizes
from 3 to 10AU.

One consequence of the existence of a population of small dark clouds is the
transient obscuration of background stars (Gerhard \& Silk 1996). Such
events are expected to be very rare with much less than 1\% of stars in any
given direction being obscured at any time.  However, even a very low rate
of obscuration events is detectable with the data from microlensing
experiments, since millions of stars have been observed for many years
(Alcock et al.~2000; Udalski et al.~2000; Derue et al.~2001; Bond et
al.~2002).  Models for the obscuration event timescale distribution
of cloud occultation events have been presented by KBS.  They
consider two hypothetical cloud populations, one in which the clouds occupy %lie within
the disk and another where they reside in the Halo.  

In this paper we present the results of an analysis of the MACHO light curve
dataset to detect light curves consistent with extinction due to passing
dark clouds. We will search for events toward stars in the disk (Galactic
bulge) and the Halo (LMC and SMC). To quantify our results we determine our
detection efficiency for the various possible dark cloud parameters in the
KBS models.

\section{\sc Observations}

The MACHO project repeatedly imaged $\sim$ 67 million stars in a total of
182 observation fields toward the Galactic bulge, the LMC and SMC, to detect
the phenomenon of microlensing (Alcock et al.~1997).  The data was taken in
$\sim$ 90,000 individual observations, each covering a total area of
$43\arcmin \times 43\arcmin$ on the sky. When field overlap is considered, the
182 fields observed cover approximately 80 square degrees.  These images
were taken between 1992 and 2000 on the Mount Stromlo and Siding Springs
Observatories' 1.3M Great Melbourne Telescope with the $8 \times 2048^{2}$ 
pixel dual-colour wide-field {\em Macho camera}.
Observations toward the Galactic bulge have exposure times of 150 seconds
while toward the LMC and SMC they have 300 and 600 seconds, respectively.
The photometry was carried out using a fixed position PSF photometry package
derivative of the DoPhot package (Schecter, Mateo \& Saha 1993) called
SoDOPHOT (Bennett 1993).  The images were taken with non-standard $B_{M}$
and $R_{M}$ bands and can be converted to the standard Kron-Cousins system
$V$ and $R$ using the photometric calibrations are given in Alcock et
al.~(1999). The dataset provides a consistent set of photometry for stars
spanning the duration of the experiment.

The observations toward the LMC and the SMC were taken all year round, while
the Galactic bulge was observed between March and October.  The median
seeing of the data set is roughly $2\arcsec$ and measurements reach stars
with V-band magnitudes between 21 and 22.  The photometric sampling
frequency varies greatly between the individual fields observed toward the
Galactic bulge and the LMC. The least sampled fields have only $\sim 50$
observations while those most sampled have $\sim 2000$ observations.
For most fields, observations are quite evenly spaced over the experiments
duration (apart from the observing gaps toward bulge fields).  However, a
few fields in the LMC and the Galactic bulge have variations in observation
frequency because of changes in the observing strategy.  Also, a small
number of the bulge fields were only observed in the last few years of the
project.

To select candidate obscuration events from the MACHO database we performed
a number of cuts to remove data which was noisy or had a low signal-to-noise
ratio. 
We selected stars with magnitudes between $13 < V < 21$ toward the
Galactic bulge and LMC and $14 < V < 22$ for the SMC.  For each of the stars
in this range we determined the median magnitude, $M_{10}$, and the standard
deviation, $\sigma_{10}$, based on the first 10\% of the data points.
However, if there were less than 500 observations in a light curve we used 
the first 50 data points.  We rejected any light curves where $\sigma_{10}$ was
larger than $0.7$ magnitudes in either passband. Altogether these cuts removed 
$\sim40\%$ of the stars observed. %(mainly LMC stars $V > 21$). 
The parameters of the analyzed dataset are given in Table \ref{tab1}.

\placetable{tab1}

\section{\sc Searching for Obscuration Events}

In this analysis we aimed to detect stellar occultations due to opaque dust
clouds and obscurations due to dark clouds with a small dust-to-gas ratio.
We also tried to detect gas cloud transits without restricting the cloud
shapes to purely spherical morphology or isotropic dust distribution.
Therefore, to find events due to the passage of dark clouds in front of
stars, we were as liberal as possible in our selection of the light curves
which might exhibit extinction effects.  

Firstly, we selected light curves that exhibited a significant drop in flux
that lasted $>10$ days.  This selection was required to remove eclipsing
binaries and data that was affected by periods of bad observing conditions.
Secondly, we required that the photometry points during the candidate
obscuration event were $>0.2$ magnitudes below the median magnitude and had
$> 4\sigma$ significance.  The uncertainty, $\sigma$, was based on the
distribution of the photometry values rather than the individual photometric
uncertainties.  As the detection of candidate obscuration events is
sensitive to the sampling rate of a target field, we also
required at least 10 points below the median in any candidate event.  To
reduce the number of long period variables detected, we only accepted light
curves with $< 50\%$ of the points below $M_{10}$, (the 10\% median value).
However, we did not require that the flux in a candidate's light curve
returned to the median value.  In this way, we hoped to retain some of the
sensitivity to dark cloud extinction events with timescales longer than the
Macho projects baseline ($\sim 8$ yrs).

\begin{figure}[ht]   
\plotone{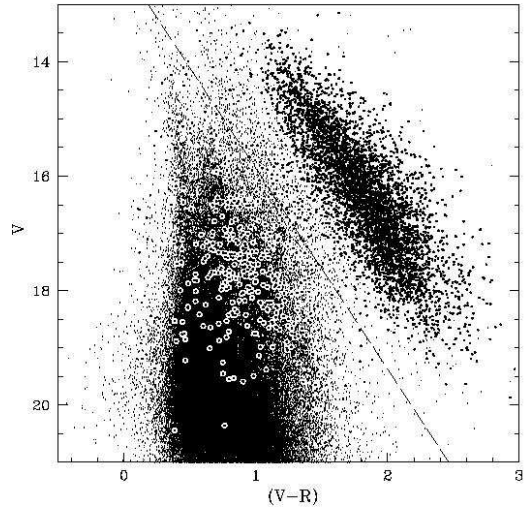}
\figcaption{The initial obscuration event candidates selected from all Galactic
bulge light curves. Stars toward the bulge fields are represented by
small dots while candidates are represented by larger ringed-dots.
The dashed-line shows the selection criteria used to remove variables from
our lists.\label{f1}}
\end{figure}

\placefigure{f1}

\begin{figure}[ht]   
\plottwo{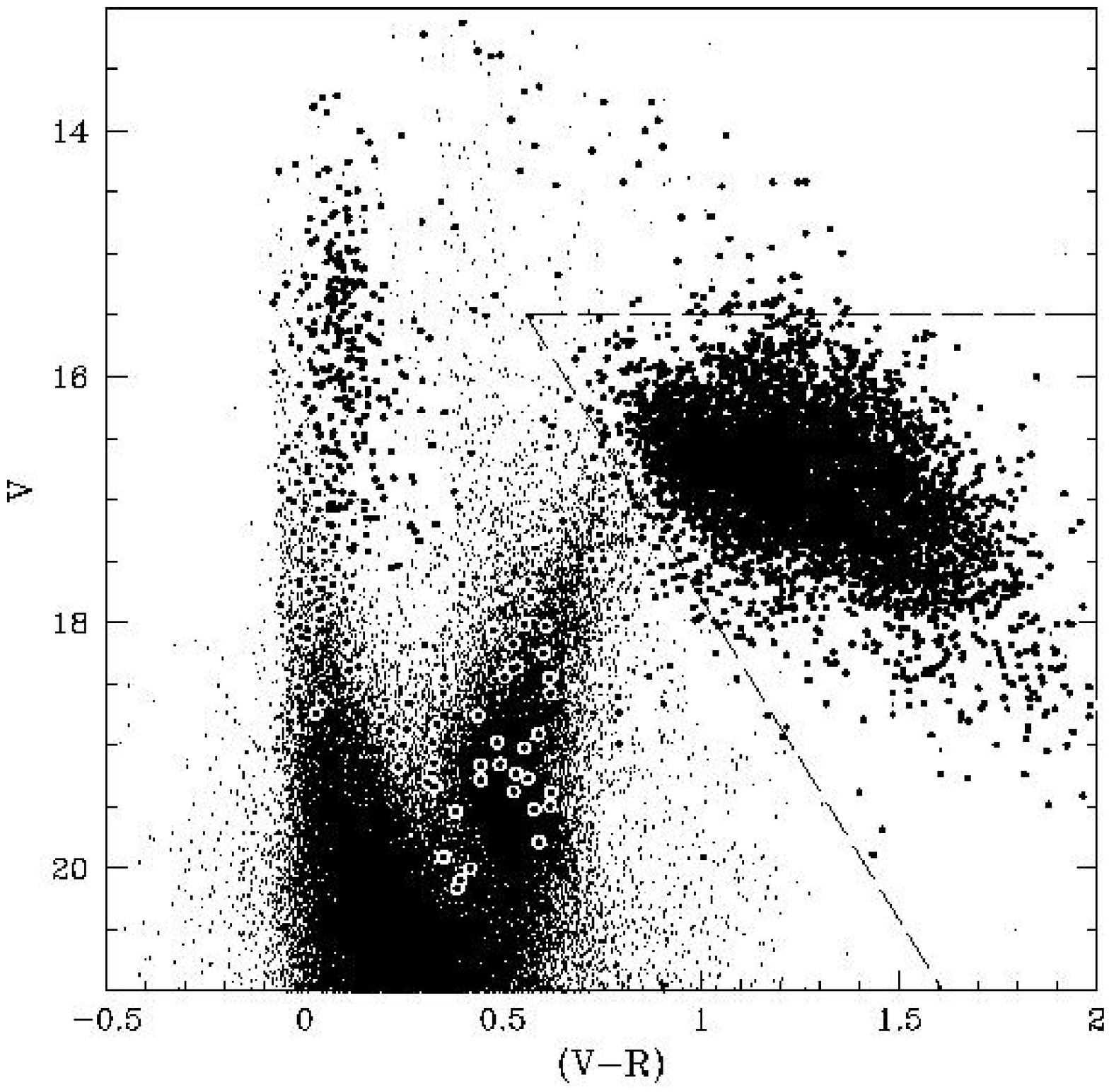}{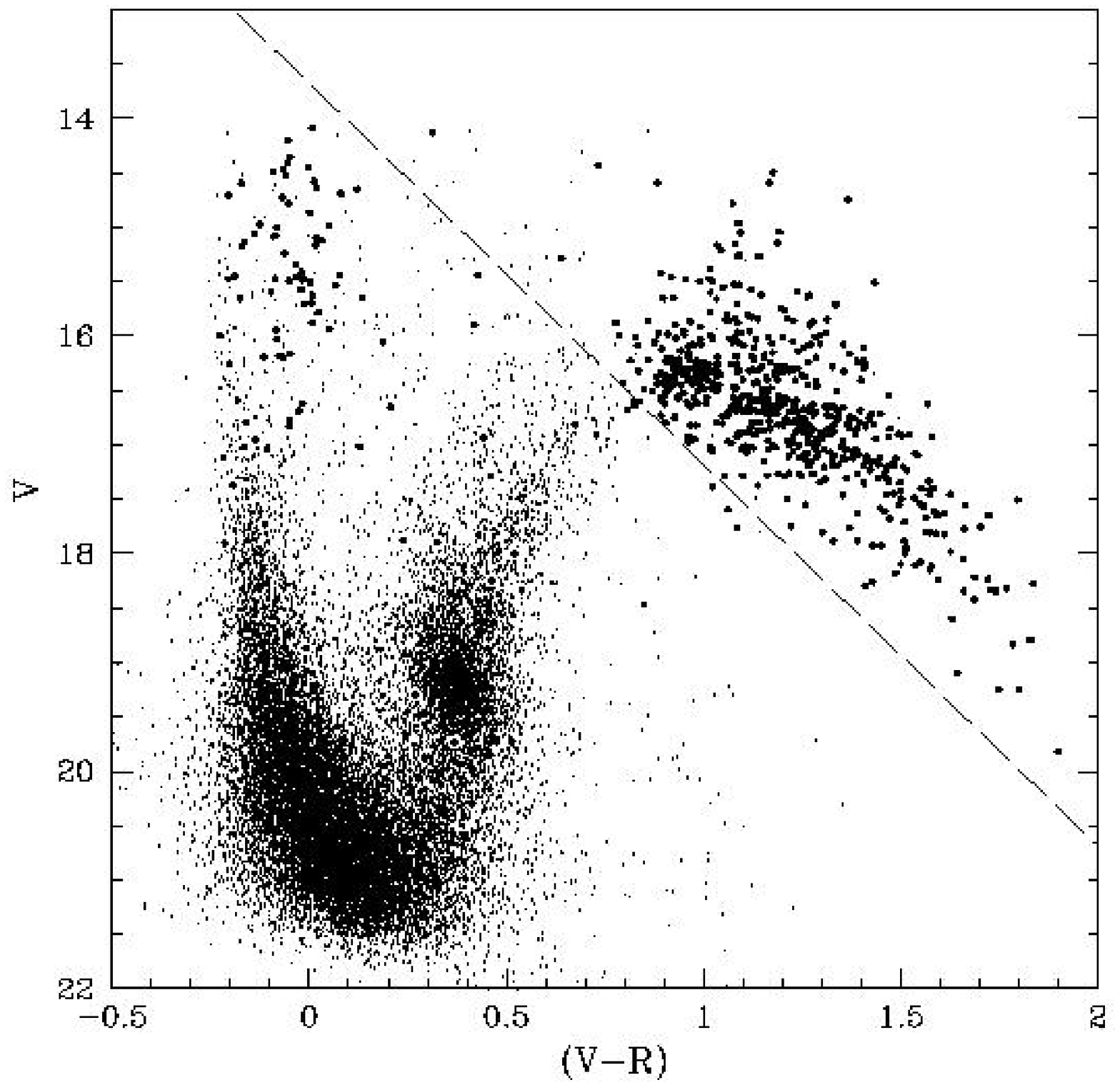} 
\figcaption{LMC and SMC dark cloud obscuration event candidates. Left panel:
  the median magnitudes for candidate light curves toward the LMC.  Right
  panel: same as the left, but for candidates toward the SMC.  The small dots
  are normal stars within the Magellanic cloud fields and larger ringed-dots are the
  candidates.  The candidates to the right of the dashed-line are deemed to
  be due to normal stellar variability.
\label{f2}}
\end{figure}

\placefigure{f2}

Applying these criteria to the MACHO project photometry data, 5284
($\sim 0.016\%$) %($\sim 0.0157\%$) 
light curves passed our initial selection toward the Galactic bulge,
6427 ($\sim 0.050\%$) %($0.0495\%$) 
toward the LMC and 678 ($\sim 0.027\%$) %($0.0273\%$) 
toward the SMC.  %Upon examination of the selected light curves, 
From examination of the candidate's light curves and their positions in
CMDs, it was clear that most of the objects passing these selections were 
long period variables. %LPVs. 
This was to be expected since these stars have large variability amplitudes,
timescales $\ga 80$ days, and red colours.  However, we intended to examine
the light curves before applying any colour selection criteria which might
bias our results.  As none of the light curves in the AGB region of the CMD
appeared consistent with dark cloud transit events, we chose to apply
empirical colour selections to the candidates toward each target direction.
In Figures \ref{f1} and \ref{f2} we present CMDs of the Galactic bulge, LMC
and SMC stars. The CMDs are composed of a $\sim 1000$ stars from each of the
182 fields observed. In each figure we have over-plotted the $M_{10}$
magnitudes of the light curves passing our primary selection as obscuration
events.  We have also plotted the colour cuts we applied to remove the AGB
variables (to the right of the dashed-line) which passed our initial
selection criteria.  As can be clearly seen, this colour selection removes
almost all the initial candidates. However, these colour cuts only remove
$\sim 1\%$ of all stars toward any of the target fields, so they do not
significantly change our total exposure.
After applying the red variable star cuts there were 372, 441 and 98 remaining
candidates toward the Galactic bulge, LMC and SMC, respectively.  Many of
these objects were also clearly variable stars.

Toward the Galactic bulge many of the candidate events had clump star
sources. Almost all of these light curves exhibit sinusoidal modulations
superimposed on a varying baseline magnitude.  The oscillation periods were
found to be between 11 and 90 days with amplitudes between 0.02 and 0.2
magnitudes.  These stars exhibited baseline magnitude changes of up to
$\sim$0.5 magnitudes. All these features are typical of chromospherically-active 
stars (Fekel, Henry, \& Eaton 2002) and can not be explained simply by a 
transiting dark cloud.
%while their counterparts in the
%Magellanic clouds appeared to have smaller values.  
Of the candidate events 155, 10, and 2 light curves (Galactic bulge, LMC,
SMC) were clearly due to chromospherically-active stars and were removed.
The variation in the number of objects toward the target fields is due to
stellar populations differences and to the fact that the photometry of clump
stars is poorer at the distances of Magellanic cloud stars.

A large number of the candidates in the Magellanic cloud fields were
found to be due to the well known population of bright, blue, $Be$ variable
stars known as bumpers (Keller et al.~2002).  These stars occupy a
restricted region of the CMD and can show long dips similar to those
expected for a cloud transit.  However, many also exhibit outbursts similar
to dwarf novae.  Such flares are not easily explainable in a cloud transit
model.  We removed an additional 89, %77 variable + 7 bad data + 5 unknown = 89
280, and 55 (Galactic bulge, LMC, SMC) exhibit either obvious flares
or multiple dips below the baseline magnitude.  Although it
is possible that a star could be eclipsed by more than one cloud within a
period of a few years, this seems extremely unlikely given the the small
number of candidates relative to the total number of stars analyzed.
%We only removed those light curves which actually exhibit
%significant ($>10\sigma$) outburst events.  
%These stars appear to be rarer toward the Galactic bulge fields because of
%stellar population differences.

Of the candidate cloud transit events 45, 18 and 1 (Galactic bulge, LMC,
SMC) were discovered to be due to high proper motion stars.  These objects
have very characteristic light curves shapes.  In such curves the dip in
magnitude is actually due to the movement of the star away from the fixed
position where photometry is performed (Alcock et al.~2001).  These light
curves show increased scatter with time and never return to the baseline
flux level. In our selection process we only removed the light curves which
exhibited increasing scatter with time to a degree significantly larger than
expected for stars of their measured magnitudes.
After removing the variable stars, etc, 83 Galactic bulge, 133 LMC,
and 36 SMC candidates remained.

If the dips observed in candidate's light curves are due to extinction
caused by dusty clouds, the obscured stars should become redder as they
fade along the extinction vector.  The exact amount of reddening should
depend on the quantity and nature of the dust present.  To test this
scenario, we assumed that the detected drop in flux was purely due to dust
extinction.  In this case the magnitude during an obscuration event should
be well represented by,

\begin{equation}
V = C + R_{VR}\Delta(V-R),
\end{equation}

\noindent
where C is the baseline flux, $\Delta(V-R)$ is the change in the star's
colour and $R_{VR} = A_{V}/E(V-R)$ is the ratio of total to selective
extinction for the bands we observed.  The $V$-band light curve of each
candidate was fitted to determine the value of $R_{VR}$.

We expect that, if the drop in flux was only due to varying amounts of dust
(producing varying degrees of extinction), the reduced $\chi^2$ of this fit
should be close to 1. However, if the drop in flux is caused by variability
which is not consistent in the two passbands the fit will be poor.
Furthermore, if the dust composition within a cloud follows standard dust
properties, the value of the extinction should be similar to that expected
for standard reddening law $R_V = A_{V} / E(B-V) = 3.1$ (Cardelli, Clayton,
\& Mathis 1989).
However, there is some evidence for variations from the standard extinction
law toward the Galactic bulge. For example, Udalski (2002) found $R_V$
values between 1.8 and 3.3 in the extinction ratio for Galactic bulge fields. %$R_{V} = A_{V} / E(B-V)$
Low $R_V$ values have also been observed by Szomoru \& Guhathakurta
(1999), whom found $R_{V} \la 2$ for a sample of Galactic cirrus clouds.
However, Clayton \& Cardelli (1988) reported that extinction values are 
typically between 2.6 and 5.5. %Higher density regions have higher Rv.
The observed differences in $R_{V}$ from the standard value have been attributed 
to variations in silicate and graphite grain size distributions 
(Kim, Martin, \& Hendry 1994, Larson et al.~2000).
To be comprehensive in our candidate event selection, we selected
light curves having fit coefficients corresponding to $R_{V}$ between 1.8
and 5.5.  Using the relative extinction for our passbands
from by Schlegel, Finkbeiner, \& Davis (1998) we
transform our results from $R_{VR}$ to $R_{V}$ 
%light curves with $R_{VR}$ values between 2.8 and 8.6. 
We adopted the same $R_{V}$ limits for the candidate light curves toward 
all targets.

\begin{figure}[ht]   
\plotone{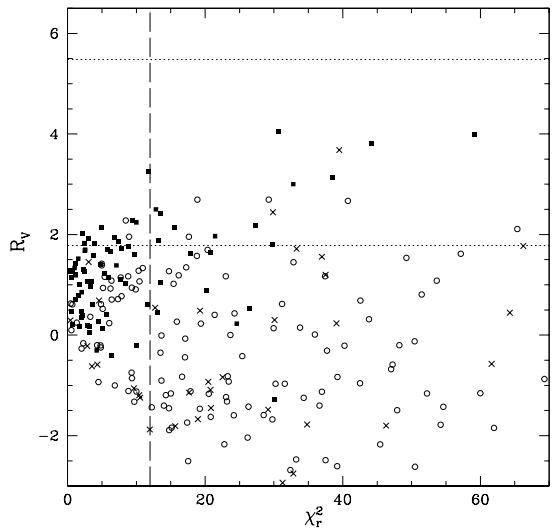}
\figcaption{Reddening fit values for candidate cloud obscuration events.
The dotted line show the expected limits if the dips on the 
light curves are due to dust reddening. The dashed line shows
maximum $\chi^{2}_{r}$ expected for the fit in the presence 
of underestimated errors and varying dust composition along the 
line-of-sight through a cloud. The filled squares show Galactic bulge 
candidate fits, while the circles and crosses present results for LMC 
and SMC candidates, respectively.\label{fx}}
\end{figure}

\placefigure{fx}

The reduced $\chi^2$ fit value is sensitive to how well the magnitude
uncertainties are determined and whether the dust composition varies within
a cloud. The flux measurement uncertainties are highly unlikely to be
incorrect by more than a factor of 3. The degree of variation in the dust
composition within a cloud is completely unknown, but it seems unlikely to
be large.  Most of the light curves are well sampled with a few hundred data
points.  Therefore, it seems unlikely that a $\chi^2_{r}$ value greater than
12 could occur even with poorly estimated photometric uncertainties and
varying dust composition within a cloud.  In Figure \ref{fx} we plot the
values of $R_{VR}$ for the MACHO V-band light curves against the reduced $\chi^2$
values of the fit.

\placetable{tab2}

Of the remaining candidates, 10 bulge, 2 LMC and 0 SMC light curves %24,10,2 no chi cut
passed our extinction criteria. In Table \ref{tab2} we present the number of
candidate obscuration events remaining after each selection criterion was
applied. From examination of their light curves it seems likely that most,
if not all, of these light curves are due to variability rather than
extinction effects. Furthermore, they do not appear to be near the location
representative of the most common stars. This suggests that they are simply
outliers among the thousands of variable stars present in the fields.

\begin{figure}[ht]   
\plotone{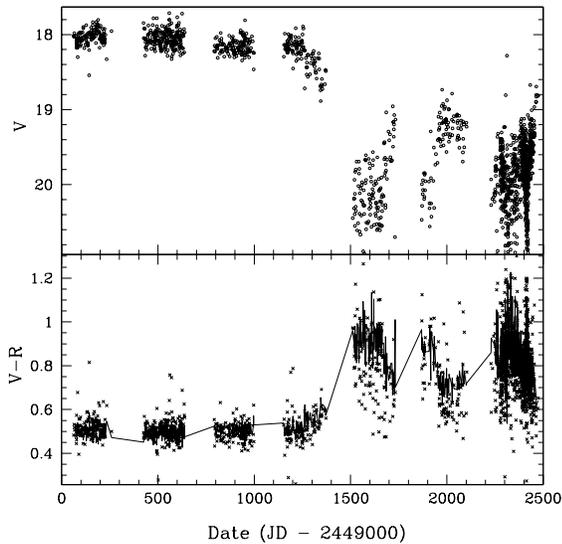}
\figcaption{The light curve for the best dark cloud transit candidate (MACHO object 
ID 119.20740.1031). In the upper panel we present the V-band light curve and in the 
lower panel the colour curve. Over-plotted on the lower panel is the fitted 
colour coefficient for this light curve.\label{fy}}
\end{figure}

\placefigure{fy}

In Figure \ref{fy} we present the light and colour curves for the Galactic
bulge candidate which appears the most like that expected for a cloud
transit event.  In this case, the object gets redder when it fades as
expected for a cloud transit.  For the standard reddening law ($R_{V} =
3.1$), we expect to measure $R_{VR} =5.2$ in our filters.  The fit of the
light curve of this candidate gives $R_{VR} = 5.1\pm0.1$.  Although the
agreement is very good, the photometry of the faint data points is very
uncertain.  Almost all of the other candidate's light curves exhibit drops
$< 1$ magnitude.  However, the Galactic bulge light curve with MACHO ID
number 120.21786.958 has a drop $> 3.5$ magnitudes and is also a good
candidate obscuration event.  If any of the the other candidates were due to
dark cloud events they would have to have a very small dust-to-gas ratio
compared to molecular clouds. The lack of a significant number of
obscuration event detections is significant, but must be quantified to
provide a useful limit on cloud models.

\section{\sc Detection Efficiencies}

To determine the significance of our result it is necessary to know how the
selections we have made will affect our efficiency of detection.  Any
obscuration event will have a start time, a timescale, a maximum extinction,
and an impact parameter.  The expected occultation timescale distribution
for viralized dark clouds with T$= 10$K and R $= 7$AU in fields toward the
Galactic bulge, LMC, and SMC, is given by KBS.  The KBS
models consider two possible dark cloud populations, a halo one and a disk
one. The halo model consists of clouds in an isothermal halo with a core
radius of 5 kpc, a local density of $\rm 0.01M_{\sun}pc^{-3}$ and a Gaussian velocity
dispersion of 156km s$^{-1}$.  The disk model consists of a sech-squared
disk with scalelength 2.5 kpc, scale height 190pc and density
$\rm 0.03M_{\sun}pc^{-3}$.  Here the Galactic rotational velocity of the clouds
and stars is taken to be 220km s$^{-1}$ with 25km s$^{-1}$ dispersion.
Although our analysis should be sensitive to a broader range of possible
cloud parameters than those of KBS, their models are well
constrained. Therefore, we will determine how efficiently dark clouds
following the KBS models would have been detected in our analysis. 
For simplicity, we assumed that the dust was isotropically distributed
within each cloud and that the obscuring clouds should not exhibit any
preference in their alignment with background stars.  We have, therefore,
selected cloud-star obscuration impact parameters from a uniform distribution 
between 0 and 1 cloud radii.
To determine our sensitivity to varying amounts of dust we assumed a uniform
distribution of the dust-to-gas fraction relative to Galactic molecular
clouds, $f$, to be between 0 and 0.1. This ratio corresponds to a visual
extinctions between 0 and 12 magnitudes (Binney \& Merrifield 1998).  For
our analysis, an extinction of 12 magnitudes corresponds to the opaque cloud
limit since even the brightest stars in our fields would become
undetectable.

For us to have detected any occultation events within our analysis it would
have to either start or finish within the observing period of the MACHO
project. Therefore, we produced a randomly distributed set of artificial
extinction events lying within the observational time frame.  For each of
the Galactic bulge, LMC and SMC datasets we produced $\sim 10$ million
artificial events. Many events were added to the same light curves in one
dataset (corresponding to a $2\arcmin \times 2\arcmin$ region) for each of
the 182 fields we analyzed.  Each of these datasets typically contain the
photometry for a few thousand stars.  Although this is only a small fraction
of the stars within a field, we believe this number was sufficient to
represent the overall properties of a field since the observational
properties will be the same throughout a field.  As there were less than 10
million stars chosen toward each target, we added a number of artificial
obscuration events to each light curve.  These extinction events were added
to both red and blue Macho light curves.  The observed drop in flux for each
light curve was calculated from the optical depth through the cloud at the
cloud-star impact parameter.  The extinction was taken to follow the
standard reddening law ($R_{V} = 3.1$).  Light curves with the artificial
cloud obscuration events were processed through the same selections as the
real data up until the the point where we removed the chromospherically
active stars.  At this point there were a few hundred objects found toward
each target in our survey. We have not attempted to parameterize and
implement the additional selections here since this would be a very complex
and somewhat contrived process.  However, we believe the selections we made
to remove the variable stars, etc, were fairly robust and would have little
effect on our calculated detection efficiency.
%note that in our analysis less than 0.002\% of the surveyed stars were
%removed by these additional selections.  this very complex process

\begin{figure}[ht]
\plotone{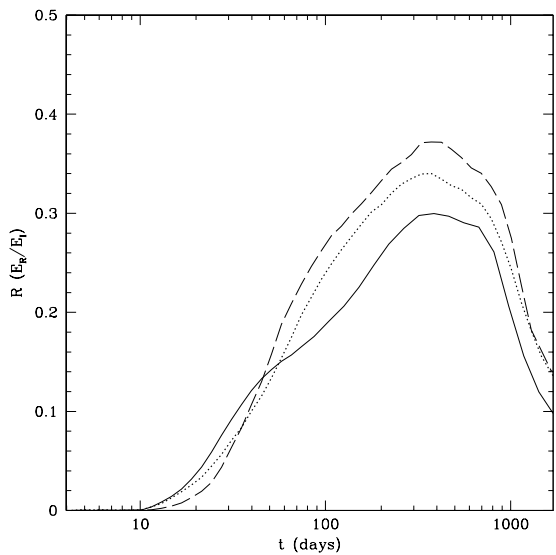}
\figcaption{Number of recovered events $R$ as a function of their input
  obscuration event timescales. Solid-lines show obscuration events
  detection efficiencies toward Galactic bulge fields, dotted-line, toward
  LMC fields, dashed-line, toward SMC fields.
\label{ft}}
\end{figure}

\placefigure{ft}

In Figure \ref{ft} we show the distribution of artificial obscuration events
recovered for the Galactic bulge, LMC and SMC. For all targets the peak
detection efficiency is around 350 days. In general, as the timescale of the
obscuration events increases, the signal-to noise ratio of even a small drop
in flux increases. However, at longer timescales the events are more likely
to be occurring at the beginning of the light curve where the baseline is
determined.  Therefore, no dip is measured relative to the median
magnitude.  At very long timescales such clouds are more likely to be
transiting throughout the observation time.  Almost all obscuration events
lasting less than ten days are not detected because of the minimum timescale
cut we imposed.

\begin{figure}[ht]
\plotone{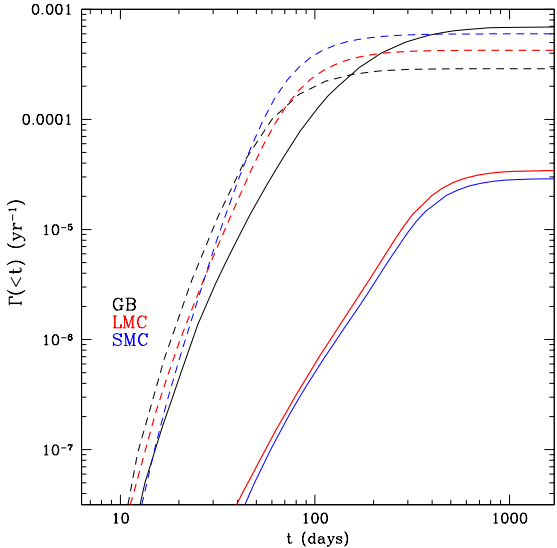}
\figcaption{The expected cumulative event rate for the KBS model of
  dark cloud occultations toward the Galactic bulge, LMC, and SMC when detection
  efficiencies are taken into account. Solid-lines represent halo clouds and the
  dashed-lines are for the disk cloud model.\label{fd}
}
\end{figure}

\placefigure{fd}

\begin{figure}[ht]
\plotone{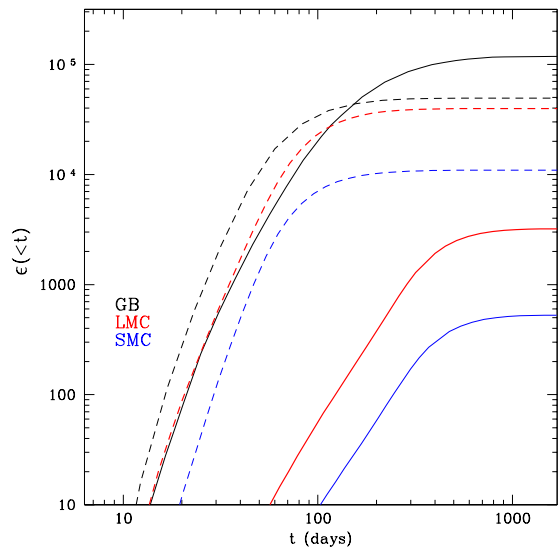}
\figcaption{Cumulative number of cloud obscuration events expected with our
  analysis for the KBS models.  The solid-lines present the halo
  cloud numbers and the dashed-lines the those for disk cloud
  models.\label{fe}
}
\end{figure}

In Figure \ref{fd} we plot the cumulative number of events expected for the
KBS models. For disk clouds, the largest number of events per star is
expected toward the Galactic bulge, while for a halo cloud population, the
largest number of events per star is expected toward the SMC.  In Figure
\ref{fe}, we plot the number of event detections expected when the stellar
exposure time is considered.  The total exposure for each target is the sum
of the number of stars monitored in a field multiplied by the number of
years it was monitored.  If dark clouds following the KBS disk model
contribute a third of disk mass density ($\rm 0.03M_{\sun}pc^{-3}$), then
$\sim 100,000$ extinction events should have been discovered toward the
Galactic bulge.  Such a large number of events could not have gone
undetected in the data.  For clouds in the KBS isothermal halo model,
the maximum number of events would have been observed toward the the
Galactic bulge rather than the LMC or SMC. This is somewhat surprising, but
is simply due to the fact that twice as many stars were observed toward the
Galactic bulge as the LMC. For this model we should have detected $\sim
50,000$ events due to cloud extinctions.

\placefigure{fe}

\subsection{Efficiencies at Low Extinction}

The number of events that we expect to detect is strongly dependent on
the amount of extinction they cause. For extinctions greater than $\sim$1
magnitude ($f \sim 0.08$) the obscuration event detection efficiency is
approximately constant for all targets. 
%At low dust concentration levels the event detection efficiency 
%increases approximately linearly between $A_V = 0.2$ and 1 magnitudes.  
To better quantify the number of obscuration events
expected if the dark clouds have very small dust concentrations ($< 1\%$
that of molecular clouds), we performed a second set of detection efficiency
simulations for extinctions less than one magnitude.

\begin{figure}[ht]
\plotone{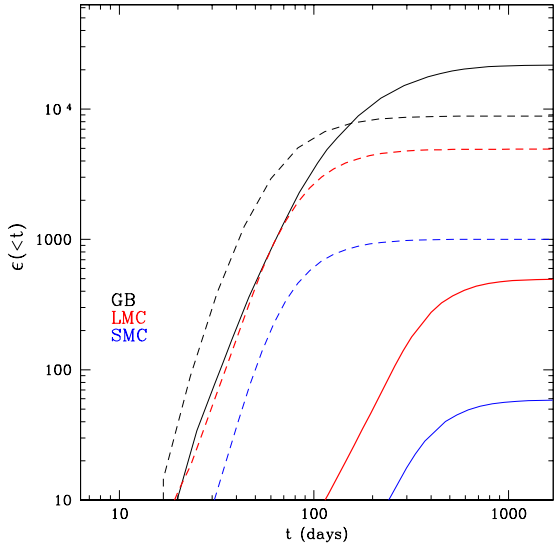}
\figcaption{Cumulative number of cloud obscuration events expected for the
  KBS model when $A_V < 1$.  The solid-lines present the halo cloud
  numbers and the dashed-lines those for disk cloud models.\label{ff}
}
\end{figure}

\placefigure{ff}

\begin{figure}[ht]
\plotone{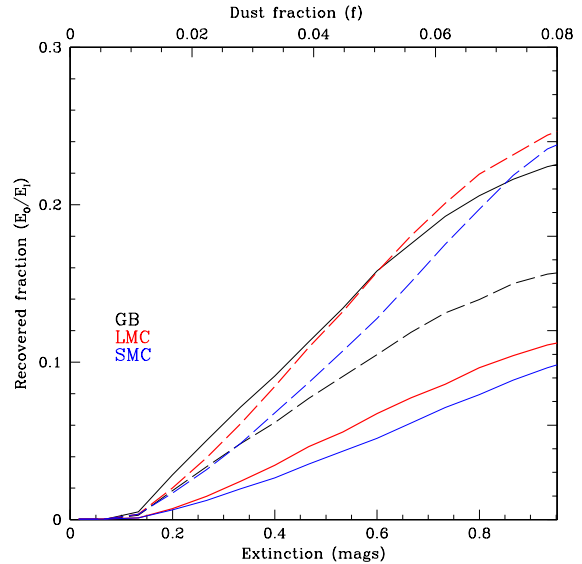}
\figcaption{Obscuration event detection efficiency number for the
  KBS model when $A_V < 1$.  The solid-lines present the halo cloud
  numbers and the dashed-lines those for disk cloud models.\label{fr}
}
\end{figure}

\placefigure{fr}

In Figure \ref{ff} we present the number of low extinction clouds we expect
to have detected in our analysis for the KBS models.  In this case we still
expect to have detected 20,000 events toward the bulge for disk clouds, or
9,000, if the clouds exist in an isothermal halo population.  In Figure
\ref{fr} we present the obscuration event detection efficiency as a function
of the extinction caused by the cloud.  The differences between the plotted
detection efficiencies are mainly due to the different timescale distributions. 
The expected number of detections only falls to zero near the 0.2 magnitudes
($f \sim 0.2\%$) because of the cut we imposed in our selection.  A
non-standard extinction law would allow a larger amount of dust to go
unnoticed. However, this would only change the result by a factor of $\sim
2$.

The dark cloud occultation event timescale is proportional to the cloud size
$R$ and the event rate is proportional to $T^{-1}R^{-1}$.  In this analysis
we have assumed the KBS model values (T$ = 10$K, $\rm M
=10^{-3}M_{\sun}$, and $R \rm = 7 AU$).  For different dark cloud models the
extinction events may appear as many short events or a small number of longer
timescale events.  For the adopted model, halo events peak in number at
around 60 days, while for disk clouds the peak is near 100 days.  From our
efficiencies we should still have been able to detect
obscuration events due to small clouds (with maximum $\sim$10 days) or due
to large clouds (with maximum number $\sim$1000 days), because the cloud
transit timescale distributions are broad. However, de Paolis (private
communication) suggests that there are models in which large halo clouds
would take 10 to 100 years to transit a background star. In such cases we
still expect to have discovered a few clouds as they began to obscure stars
(provided they produced more than 0.2 magnitudes of visual extinction).
Our results do not show any trace of a low-mass dark cloud population.

\section{\sc Discussion}

In our analysis we have attempted to maximize our efficiency for
detecting dark clouds obscuration events over a range of characteristic
timescales and degrees extinctions.
Our selection process should have allowed us to detect extinctions due to
passing clouds even if they were non-spherical or had an internal dust
distribution which was slightly anisotropic.  However, we do not find any
evidence for a population of dark clouds in either the disk or halo of
our Galaxy.  We can strongly rule out the existence of a population of dark
($A_{V} > 0.2$) clouds with the parameters of the KBS disk or halo model.

There are a number of possible ways a population of gaseous clouds may have
been missed in this analysis.  For instance, such clouds could have a very
small dust content and therefore be almost completely transparent. In such
cases gas lensing events can still occur.  The mass and radius limits for
possible clouds from Rafikov and Draine (2001) are approximately $\rm 10^{-4} \la M \la
2\times10^{-2}M_{\sun}$, $R < 200$AU (see their Figure 9). 
%do not exclude the entire expected mass and radius range for a dark 
%cloud population (see their Figure 9).  However, 
While our results seem to rule out the possibility that a population of dark
clouds exist with masses from $10^{-4}$ to $\rm 10^{-2}M_{\sun}$.  Together
these results exclude the possibility of a dusty cloud population with a
polytropic index of 1.5.  Whether the proposed population of gaseous clouds
are dusty is not certain. However, as we mentioned earlier, if the SCUBA
sources of Lawrence (2001) are due to dark clouds, they must contain some
dust to show continuous emission in the sub-mm band.  The existence of dusty
molecular clumps is also useful for explaining the EGRET background
$\gamma$-ray emission results (Kalbera et al.~1999).  It is possible that
the dust in the clouds would not be readily observed from stellar obscuration
events. For instance the dust may sediment to the core of the cloud
(Wardle \& Walker 1999). Alternatively, the clouds may be clumped into large
dark clusters which exist only in regions far from the Galactic disk (
Wasserman \& Salpeter 1994, de Paolis et al.~1995, 1998).

We would like to thank Kim Griest and Francesco de Paolis for encouraging us
to perform this work.  This work was performed under the auspices of the
U.S.~Department of Energy National Nuclear Security Administration by the
University of California, Lawrence Livermore National Laboratory under
contract W-7405-Eng-48.

%\include{table1}
% TABLE1.TEX
\begin{deluxetable}{crrcr}
\tablecaption{Observational Data.\label{tab1}}
\footnotesize
%\small
\tablewidth{0pt}
\tablehead{\colhead{Target}  &  \colhead{Obs} & \colhead{Stars} & \colhead{Fields} & \colhead{Exposure}
}
\startdata
Bulge &  32856   & $3.32 \times 10^{7}$ & 94  & $1.71 \times 10^{8}$\\
LMC   &  45270   & $1.29 \times 10^{7}$ & 82  & $9.38 \times 10^{7}$\\
SMC   &   5975   & $2.48 \times 10^{6}$ & 6 &  $1.83\times 10^{7}$\\
\hline
Total  & 84101 & $4.85 \times 10^{7}$  & 182  &  $2.83 \times 10^{8}$\\
\enddata
\tablecomments{Col. (5), the exposure is in star years.}
\end{deluxetable}

% TABLE2.TEX
\begin{deluxetable}{crrrrrc}
\tablecaption{Candidate Obscuration Events.\label{tab2}}
\footnotesize
%\small
\tablewidth{0pt}
\tablehead{\colhead{Target} & \colhead{$\rm C_{Init}$} & \colhead{$\rm C_{col}$} & 
\colhead{$\rm C_{CA}$} & \colhead{$\rm C_{var}$} & \colhead{$\rm C_{HPM}$} & \colhead{$\rm C_{R_{V\!R},\chi^2_{r}}$} 
}
\startdata
Bulge &  5284  & 372 & 217 &  128 &  83 & 10\\
LMC   &  6427  & 441 & 431 &  151 & 133 & 2 \\
SMC   &   678  &  98 &  96 &   37 &  36 & 0 \\
\enddata
\tablecomments{Candidates present toward the various targets after apply
various selections:
Col.~(2), initial candidates. 
Col.~(3), colour cuts are used to remove bright red variables. 
Col.~(4), removing chromospherically-active stars. 
Col.~(5), removing Be and semi-periodic variable stars.
Col.~(5), removing high proper motion stars. 
Col.~(6), removing objects with fits $\chi^{2}_{r} > 12$ or $R_{V\!R} < 2.8$.
}
\end{deluxetable}

\end{document}